\documentclass[sigconf]{acmart}

\usepackage{booktabs} 
\usepackage{epsfig}
\usepackage{amssymb}
\usepackage{amsmath}
\usepackage{tabularx}
\usepackage{tabularx}
\usepackage{tabularx,ragged2e}
\newcolumntype{C}{>{\Centering\arraybackslash}X} 
\usepackage{amsfonts}
\usepackage[utf8]{inputenc}
\usepackage[T1]{fontenc}
\usepackage{tabu}
\usepackage{hyperref}
\usepackage{multirow}
\usepackage{makecell}
\usepackage{ctable}
\usepackage{capt-of}
\usepackage{varwidth}
\usepackage{balance}

\copyrightyear{2018}
\acmYear{2018}
\setcopyright{acmcopyright}
\acmConference{Arxiv}
\acmDOI{ }
\acmISBN{ }

\hypersetup{draft}
\begin{document}
\title{Intrinsic and Extrinsic Motivation Modeling Essential for Multi-Modal Health Recommender Systems}


\author{Nitish Nag}
\affiliation{%
  \institution{University of California, Irvine}
  \city{Irvine} 
  \state{California} 
}
\email{nagn@uci.edu}





\author{Mathias Lux}
\affiliation{%
  \institution{Alpen-Adria-Universit\"at Klagenfurt}
  \city{Klagenfurt} 
  \state{Austria} 
}
\email{mlux@itec.aau.at}

\author{Ramesh Jain}
\affiliation{%
  \institution{University of California, Irvine}
  \city{Irvine} 
  \state{California} 
}
\email{jain@ics.uci.edu}

\begin{abstract}
Managing health lays the core foundation to enabling quality life experiences. Modern computer science research, and especially the field of recommender systems, has enhanced the quality of experiences in fields such as entertainment, shopping, and advertising; yet lags in the health domain. We are developing an approach to leverage multimedia for human health based on motivation modeling and recommendation of actions. Health is primarily a product of our everyday lifestyle actions, yet we have minimal health guidance on making everyday choices. Recommendations are the key to modern content consumption and decisions. Furthermore, long-term engagement with recommender systems is key for true effectiveness. Distinguishing intrinsic and extrinsic motivations from multi-modal data is key to provide recommendations that primarily fuel the intrinsic intentions, while using extrinsic motivation to further support intrinsic motivation. This understanding builds the foundation of sustainable behavioral adaptation for optimal personalized lifestyle health benefits.
\end{abstract}

%
%





\maketitle

\section{Introduction}
Health is essentially a product of our genome and lifestyle \cite{topol2016patient}. Lifestyle is the primary controllable aspect of our health. Although largely preventable, lifestyle associated diseases continue to lead the rapid rise of non-communicable diseases across the globe~\cite{world2011global}. Producing changes in routine lifestyle habits is a tremendous psychological hurdle. Media platforms such as food analysis from images, accelerometer based activity sensors, and life logging~\cite{gurrin2014lifelogging} gather data about lifestyle factors, but largely remain as isolated efforts to collect, manage and provide ad-hoc retrieval. The next step will be to use this data to give actionable decision recommendations to empower individuals.

\begin{figure}[t]
\centering
\includegraphics[width=.87\columnwidth]{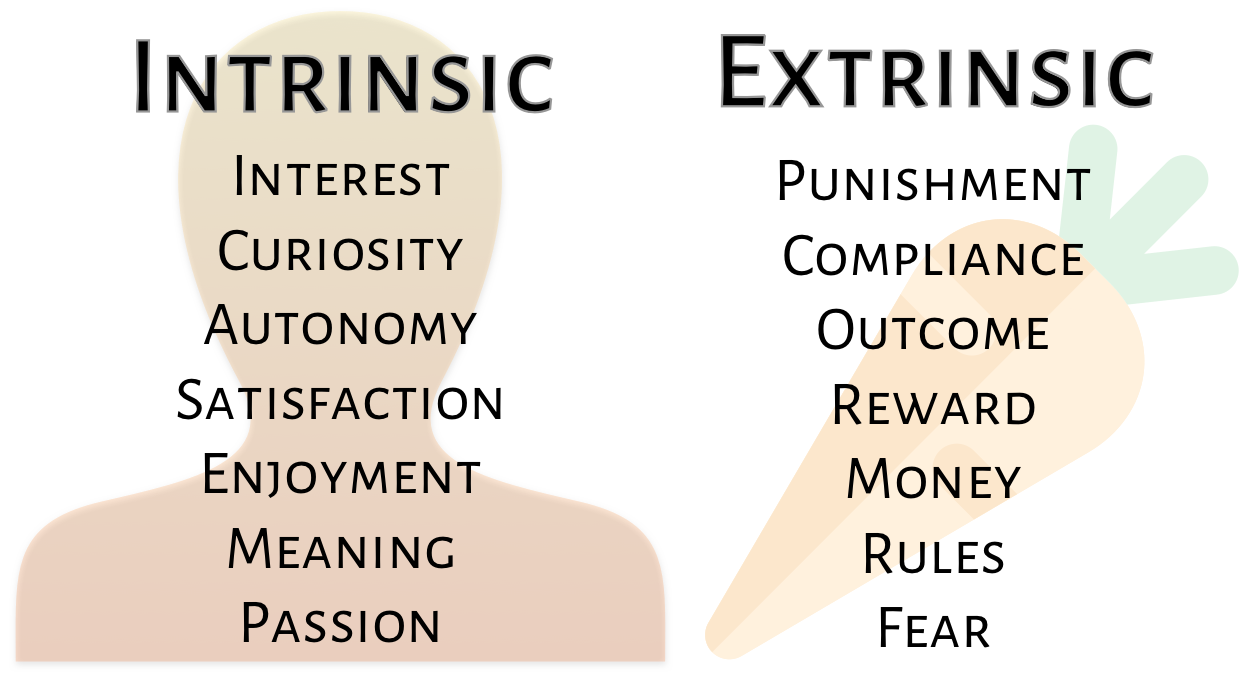}
\caption{Intrinsic motivation relies on the user enjoying the task due to a sense of personal satisfaction. Materialistic rewards for positive reinforcement and punishment for negative reinforcement are examples of extrinsic motivations, which are applied to the user from the outside world.}
~\label{fig:vs}
\vspace{-8mm}
\end{figure}

A key concept in effective lifestyle recommendation will be the distinction between intrinsic and extrinsic motivation~\cite{benabou2003intrinsic}, cp. Fig.~\ref{fig:vs}. Intrinsic motivation arises from internal enjoyment or satisfaction in doing a certain task, while extrinsic motivation refers to motivating factors from the outside world such as rewards or punishment. While both can lead to lifestyle activities that benefit health, intrinsic motivation drives long term user engagement. Self-driven healthy activities and conscious choices towards healthy behavior are only sustainable and effective in long term developments. Extrinsic motivation, like recommendations from medical doctors or therapists, social pressure through groups of likewise minded people, or an mobile phone app or fitness tracker device beeping users into action, on the other hand is a necessary part, but should only support intrinsic motivation. Intrinsic motivation is highly personalized, just like health and the reaction of the human body to different activities.

Multi-modal data can be used to learn if a user is executing an action from intrinsic motivation or from external factors. For instance, if a user repeats an activity without prompting and without influence in their social media, this gives us an increased weighting that the motivation is intrinsic. If we detect themes and concepts in their social media consumption, and the activity is done when a friend prompts them, we can consider this as more weighting towards the extrinsic motivation.

We identify two potential intuitive groups of users. These are just intuitive starting points for our position. There is no formal distinction or definition available without further research. Members both groups can be highly intrinsically or extrinsically motivated, with intrinsic motivation classically waning when immediate value is not perceived:

\textbf{Active Lifestyle Users} Sports and healthy lifestyles are fashionable for this group. This is expected from social pressure, advertising, fashion, or from intrinsic passion towards the same sports and health lifestyle.

\textbf{Health Conscious Users} Health-related activities foster long term benefits, better life, and ultimately allow this group to live a longer and more comfortable life. Delaying onset of disease is a large priority.

\section{Current State \& Models}
Considering the context for research we have right now, the trend of vast collection and data storage, mobile devices, wearable sensors, life logging, map applications, and fitness apps provide raw data and event streams. When merged and synchronized, event streams form a personicle, a personal chronicle of life events \cite{jain2014objective}, to produce a stream of time-indexed events, e.g., high fat meal eaten at 3pm Monday / 40 minutes of exercise at 2pm on Saturday etc.

Based on personicles, domain knowledge, and external data sources, like nutrition facts of meals, dietary plans, medical cases, exercise plans etc., a recommender system can suggest the next action. Important dimensions to consider are:

\textbf{Ability} encompasses the resource availability in performing the suggested behavior. This factor is an interplay of the individual's environment and their intrinsic capability, and must ensure those constraints are met. Examples are allergies, that constrain the suggestions for meals, or physical disabilities that prohibit physical abilities.

\textbf{Triggers} are instances of events that lead to accomplishment of a task~\cite{Fogg2009}. These can be optimized by using the current spatio-temporal context of the user to filter the list of recommended actionable items.
For example, a notification about healthy food options when the person is hungry.

\textbf{Motivation} captures the individual's willingness to follow through with a suggestion. The list of recommendations are ordered on the basis of the models of intrinsic preferences of the user.

Combining the right triggers with context, we can maximize the effectiveness of sustainable positive health recommendations. Over time, the recommender system should adapt to the user and identify triggers that are needed to support intrinsic motivation. For example exercising everyday should become habit rather than a constant suggestion from the recommendation engine.


%
%
%

\section{Challenges}
To the best of our knowledge the distinction between intrinsic and extrinsic motivation in recommendation engines is a novel approach. It will lead to a change in interaction from a system that helps people to find their path to a healthy life to a support system for diversification and adaptation of healthy activities.
%
The implementation of such a system poses a lot of questions including which type of recommendations and triggers it should provide and how we can put a human in the loop, which is advisable for health applications.
However, considering the novelty of the distinction between intrinsic and extrinsic motivation in health recommendation, we primarily identify the following three new challenges:

\textbf{Modeling user health motivations.} We need to build a model of the user's motivations and preferences. By using this, we can start to exploit what the user finds most interesting and especially intrinsically motivating. At the same, we need to see which recommendations that impact health stoke the intrinsic motivation.

\textbf{Recommendations that foster intrinsic motivation.} Given we know what the user is likely to find intrinsically motivating, we start to recommend the healthiest choices in a given context that uses encoded knowledge of health professionals to guide users. A simple abstraction example would be as follows: If a user is recommended eating fresh fruit on day 1, on following days the user begins enjoy fresh fruit without prompting.

\textbf{Adaptation and tracking over time and context.} Moving from extrinsic to intrinsic motivation is a long term goal. The recommender system needs to support that and adapt to the new state. If users loose their intrinsic motivation the recommender can provide a safety net to help users not to bounce back into unhealthy routines. This requires both tracking and recommendations to be synchronized.

\section{Next Steps}
We need to build a model for intrinsic and extrinsic motivation of users applicable to such a system. Sports psychology provides a good starting point for detailed investigation. There are two immediate use cases we pose as examples. One is healthy nutrition. This is promising as laws already regulates information about nutritional facts must be printed on menus and packaged foods. Moreover, in the USA Food away from home's share of total food expenditures continue to rise past 50.1\% since 2014, surpassing at-home food sales for the first time in recorded history~\cite{adfood2017}. Nutrition facts pave the way for quantitatively ranked suggestions for their meals based on past nutrition and dietary expert knowledge. 
A second opportunity is therapy supported by self-directed exercise, ie. after major surgery. Therapists provide training plans for the patient to exercise on their own responsibility. A recommender system can adapt the training plan to the users current state and can provide triggers for exercising. Therapists in the loop observe and discuss recovery and health improvement.

We strongly believe that a system according to our vision has multiple benefits: First, it gives direct feedback to quantitatively understand personal health status. Second, it empowers patient participation in health and driving user engagement. Third, it can inform the individual of risks/benefits regarding their choices. Ultimately, prevention of chronic disease allows the individual the highest quality of life while benefiting society as a whole by reducing the overall costs for health care. Focus on keeping individuals healthy instead of treating diseases and symptoms will be the next evolution in human health recommendation systems.



\section{Acknowledgments}
This research is partially funded by the National Institute of Health (NIH, United States of America) as part of the Medical Scientist Training Program (MSTP) and the Cardiovascular Applied Research and Entrepreneurship (CARE) grant under \#T32GM008620-15. Additionally funding is provided by the UC Irvine Donald Bren School of Information and Computer Science.


\bibliographystyle{ACM-Reference-Format}
\bibliography{refs}



\bibliographystyle{ACM-Reference-Format}
\balance
\bibliography{refs}

\end{document}